# Commissioning of CMS and early standard model measurements with jets, missing transverse energy and photons at the LHC

Tim Christiansen
for the ATLAS and CMS Collaborations
*CERN, CH-1211 Geneva 23, Switzerland*

We report on the status and history of the CMS commissioning, together with selected results from cosmic-muon data. The second part focuses on strategies for optimizing the reconstruction of jets, missing transverse energy and photons for early standard model measurements at ATLAS and CMS with the first collision data from the Large Hadron Collider at CERN.

## 1  CMS Commissioning

With the first collisions from the LHC expected soon, the CMS experiment [1] at CERN has entered the final stage of commissioning. Nearly all of the detectors have been installed in the experiment and the last heavy structure of CMS was lowered into the experimental hall in January 2008 (see photograph in Fig. 1). The CMS collaboration has launched over the past years a series of combined data-taking exercises, so-called Global Runs, with increasing scope and complexity. More and more components have been integrated with the trigger and DAQ systems, and data from cosmic muons as well as high-rate random triggers have been used to prove readiness for LHC collisions.

One of the first highlights in the commissioning of the CMS detector was the "Magnet Test and Cosmic Challenge" (MTCC) [2], which took place in 2006 when the CMS detector was operated in the assembly hall on the surface. The superconducting magnet of CMS required testing before lowering, providing a unique opportunity to operate all the subdetectors and sub-systems together and to take data with cosmic-ray muons. The participating systems included a 60° sector of the muon system, comprising gas detectors like the drift tubes (DTs), Cathode Strip Chambers (CSCs) and Resistive Plate Chambers (RPCs), both in the barrel part and endcaps of CMS. The tracking system comprised elements of the Silicon-Strip Tracker, and parts of the Electromagnetic (ECAL) and Hadron Calorimeter (HCAL) detected energy depositions of the traversing muons. In the second phase of the MTCC, the ECAL modules and the tracker elements were replaced by a specially designed mapping device to measure the three components of the magnetic field in the volume of the inner detectors with high precision, while HCAL and the muon systems continued to record cosmic-ray data.

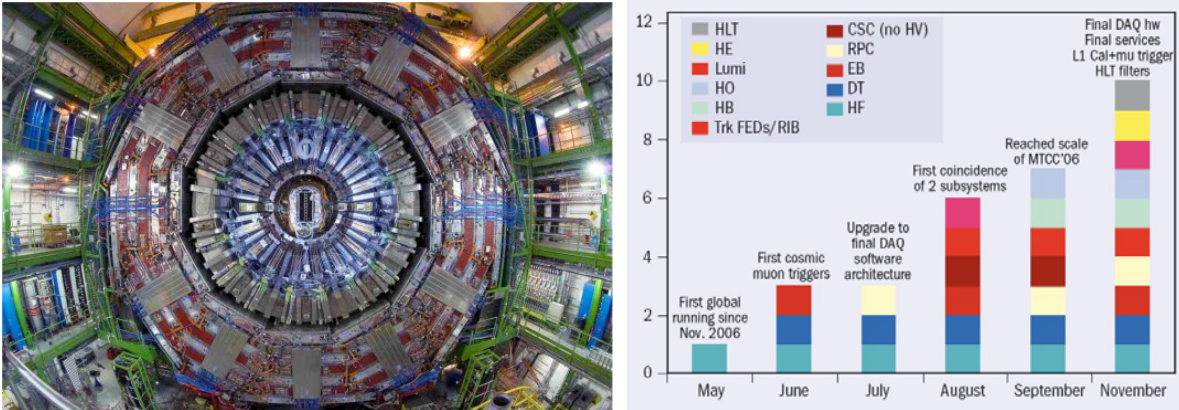

Figure 1: The CMS barrel detector in the experimental hall. The increasing number of subsystems integrated in global data taking over 2007 (right): HLT: High-Level Trigger, HE: Endcap Hadron Calorimeter, Lumi: Luminosity system, HO: Outer Hadron Calorimeter, HB: Barrel Hadron Calorimeter, Trk FEDs/RIB: Tracker frontend drivers/"rod-in-a-box", CSC: Cathode Strip Chambers, RPC: Resistive Plate Chambers, EB: Barrel Electromagnetic Calorimeter, DT: Drift Tubes, HF: Forward Hadron Calorimeter. The diagram is for illustration of the progress only; the scale is arbitrary and non-linear.

Only six months after the conclusion of the MTCC, global data taking was resumed in the first of a series of Global Runs, this time in the underground experimental hall. With the end of 2007, CMS has recorded data from synchronized cosmic triggers from parts of all trigger detectors (CSCs, DTs, RPCs, ECAL and HCAL). The diagram on the right of Fig. 2 illustrates the progress made in the integration of subdetectors and subsystems in various Global Runs in 2007. One of the many results from these exercises is the confirmation of a single-hit resolution $\Delta x < 280$ μm along $\phi$ [a] of the Muon Drift Tubes. One of the next goals before the LHC startup is the integration of the central tracker in the Global Runs.

## 2 Early Standard Model Measurements with Jets, Missing Transverse Energy and Photons

Already the first 10 to 100 $pb^{-1}$ of recorded data at the LHC experiments will allow QCD measurements with minimum-bias events and will open the window to the Z- and W-boson and top-quark production. However, it is natural to expect that the first ~ 100 $pb^{-1}$ of integrated luminosity will first be used to test and improve the understanding of the detector response and to calibrate and measure the performance of the physics objects used in the various searches and measurements to follow. Therefore, this article focuses on the strategies for the "commissioning" of the physics objects with the first data, using the standard model processes accessible with the recorded data set. Details about the longer-term physics plans of ATLAS and CMS can be found in the references [1,3].

Physicists from ATLAS and CMS are taking care that the experiments will make the best use of the early data to align and calibrate the detectors. Analyses are prepared to study and optimize the performance of the triggers and of physics objects. The strategies range from transverse-momentum balancing to exploiting well-known SM signatures, such as *tag-and-probe* with leptonic Z-boson decays. As the field is too large to cover in this article all the aspects of measurements with jets, electrons, photons and missing transverse energy, we report on a few selected recent studies from ATLAS and CMS with a focus on the reconstruction performance and strategies.

---

[a] $\phi$ is the azimuth angle of the CMS coordinate system with the $z$ axis pointing in the direction of the beam and the origin in the center of the detector. $\eta$ denotes the pseudorapidity $\eta = -\ln[\tan(\theta/2)]$ where $\theta$ is the polar angle of the CMS coordinate system.

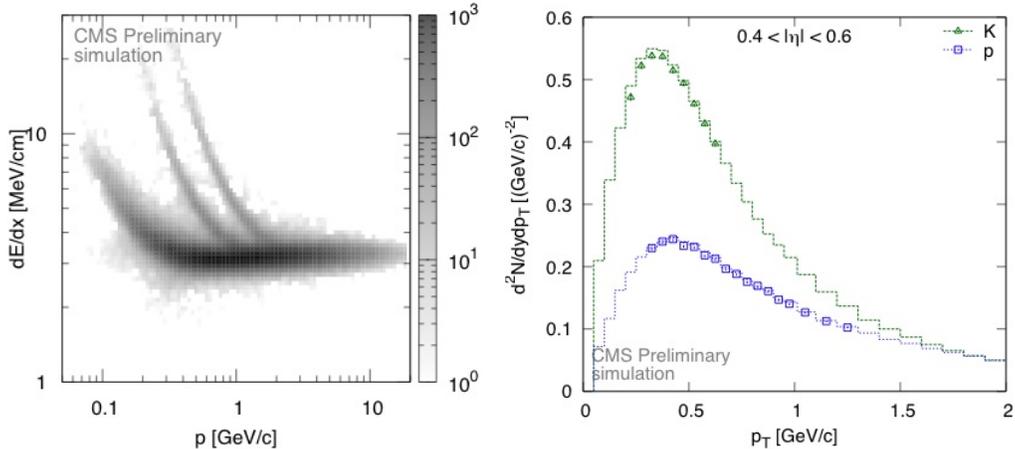

Figure 2: Left: Energy loss *dE/dx* spectra for charged pions, kaons (K) and protons (p) in the Silicon Tracker (Pixels and Strips combined) of CMS. The scatter plot clearly shows the three separate bands for charged pions, kaons and protons (from left to right). Right: Charged-hadron spectra $dN/dydp_T$ of protons and kaons as a function of transverse momentum. The "data" statistics correspond to approximately one month of data taking with 1 Hz allocated bandwidth from minimum- or zero-bias events.

2.1 Charged-Hadron Spectra

One of the first measurements to be performed at the LHC aims at the understanding of the spectrum of charged hadrons. These spectra have never been explored in hadron collisions at such high energies (collision energy > 2 TeV) and they are an important tool for the calibration and understanding of the detector response. Recent studies at CMS show that it is feasible to distinguish charged pions, kaons and protons with momenta up to 2 GeV/*c* and individually measure their spectra (see Fig. 2) [4]. Cross-sections and differential yields of charged particles (unidentified or identified pions, kaons and protons), produced in inelastic proton-proton collisions at a center-of mass energy of 14 TeV, can be measured with good precision with the CMS Pixel vertex detector and tracker system.

2.2 Underlying Event

An important ingredient for the correct simulation of standard model processes at LHC is the understanding of the "underlying event", consisting of the "beam-beam remnants" (a soft component coming from the break-up of the two beam hadrons). Furthermore, it is sensitive to test multiple-parton interaction (MIP) tunes of QCD [5]. Technically, it is impossible to fully separate the underlying event from the hard scattering process, however, the observation of charged particles in the region "perpendicular" to the leading jet ($60° < |\Delta\phi_j| < 120°$ in the transverse projection) in QCD di-jet events can be used to distinguish between various MIP tunes that have been considered [10]: DW [6], DWT [7], S0 [8] and Herwig [9] (see Fig. 3).

2.3 Jets

Up to now, the analyses of the two experiments reconstruct jets mainly with variations of the iterative cone algorithm [11] that forms jets from energy depositions in calorimeter towers in a cone of fixed size $\Delta R = (\Delta\eta^2 + \Delta\phi^2)^{-2}$. Calorimeter towers are the energy sums measured in the various depths of the calorimeter along *r* at fixed $\eta$ and $\phi$. Even though iterative cone algorithms are well established and fast, which make them particularly useful for triggering, new algorithms have been studied to improve infrared- and collinear-safety. Among those alternative algorithms are the *Seedless Infrared-Safe Cone* [12] and Fast-$k_T$ [13] algorithms.

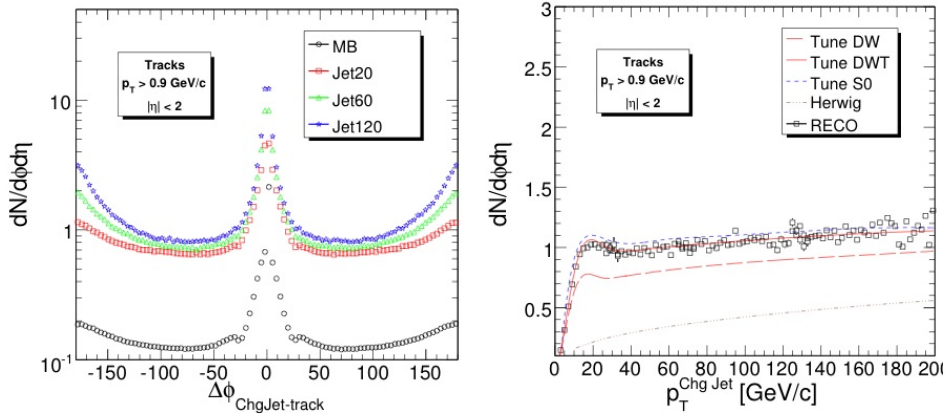

Figure 3: Densities *dN/dηdφ* of charged particle multiplicity as a function of the azimuthal distance to the leading charged jet direction $\Delta\phi$ (left). The "transverse" region used for the measurement of the underlying event is defined as $60° < |\Delta\phi| < 120°$. The data points are shown for different trigger conditions (minimum-bias (MB) or single-jet calorimetric $p_T$ thresholds), sensitive to different ranges of jet momenta. Right: Densities *dN/dηdφ* for tracks in the transverse region with $p_T > 0.9$ GeV/*c*, as a function of the transverse momentum of the leading charged jet. The simulated "data" (open squares) correspond to an integrated luminosity of 100 pb$^{-1}$.

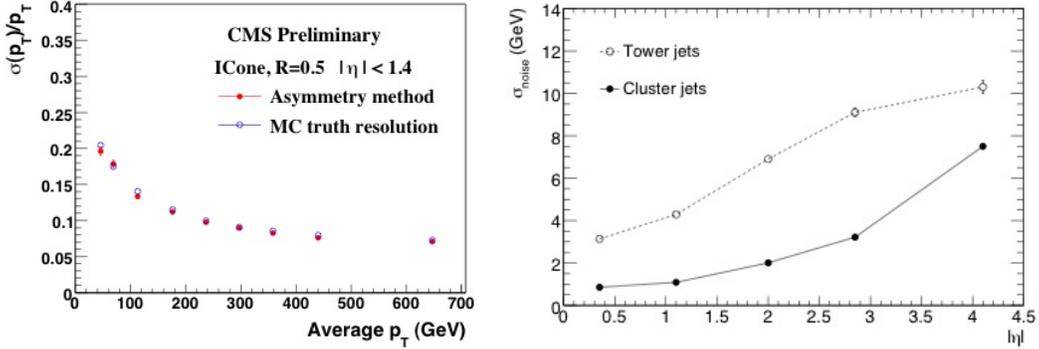

Figure 4: Jet energy resolution from a data-driven correction, based on a study of transverse-energy asymmetry in di-jet events in CMS (left). ATLAS study of the noise contribution to $R = 0.7$ cone jets made from 3D clusters compared to *standard* jets based on calorimeter towers in QCD di-jet events (right).

Rather than relying on the Monte Carlo simulation, the jet energy corrections can be extracted from transverse-energy asymmetry measured in di-jet events, a technique developed at the Tevatron [14]. In 2→2 events, transverse momenta of two jets are equal. This property can be used to scale the jet transverse momentum ($p_T$) at a given $\eta$ to a jet $p_T$ in a reference $\eta$ region, in order to correct for the variation of the jet energy response as a function of $\eta$ based on data. However, as most events in the QCD di-jet samples have more than two reconstructed jets, which primarily originate from soft radiation and which degrade the momentum balance in the transverse plane between the two leading jets, the resolution obtained from the asymmetry is studied as a function of the maximum $p_T$ of the third jet. The final resolution is then extracted from the extrapolation of the $p_T$ of the third jet in the limit $p_T(3^{rd}\ jet) \to 0$. The left plot in Fig. 4 shows how well this data-driven method compares to Monte-Carlo *truth* as studied by CMS.

It has been shown for the ATLAS experiment, that the noise contribution in jets can be reduced when using topological cell clusters as inputs to the jet reconstruction rather than the total deposited energies in the calorimeter towers. The improvement from this attempt to reconstruct three-dimensional energy depositions is shown in Fig. 4 (right) [15]. As a result of the noise reduction shown in this figure, the angular and energy resolution and the jet efficiency is expected to improve for relatively low-momentum jets (below $p_T = 40$ GeV/*c*).

It has also been shown that a clear signal of top-quark pair events, where one W boson from the top-quark decay produces a lepton (electron or muon) and a neutrino, while the other W boson decays into two quarks, can be extracted from the first 100 pb$^{-1}$ without the use of missing transverse energy or b-quark flavor tagging [16]. With the help of the W-mass constraint, such a sample will be exploited to improve the jet-energy scale uncertainty and to study jet-flavor tagging.

2.4 Missing Transverse Energy

The missing transverse energy is one of the most complex physics objects at hadron collider experiments, because it combines information from all sub-detectors and requires homogeneous calorimeter coverage in the pseudorapidity region of $\eta < 5$. A good understanding of the transverse missing energy is important, as it is present in signatures of physics beyond the standard model, but also required for the reconstruction of e.g. leptonically decaying W bosons. Various contributions of *fake* missing transverse energy can degrade the performance: machine background from the accelerator and beam-gas interactions, noisy or dead calorimeter cells or regions, non-linearity in the hadronic response and finite energy resolution. Figure 5 illustrates an example on how to make analyses less sensitive to fake missing transverse energy: ATLAS has recently studied the reduction of the fake contribution to missing transverse energy from muons escaping the fiducial region of the detector by requiring angular separation of the direction of the missing transverse energy and high-momentum jets [15].

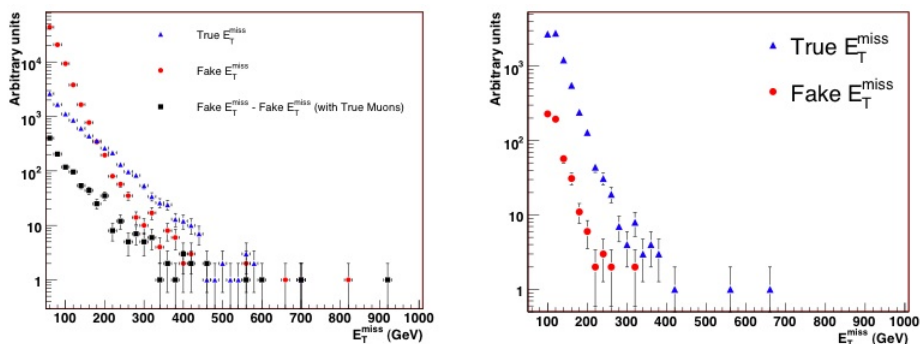

Figure 5: Distribution of missing transverse energy, fake missing transverse energy and fake missing transverse energy from escaped muons for QCD di-jet events with leading-jet transverse energies between 560 and 1120 GeV. Right: after applying an isolation cut on the azimuth of the reconstructed missing transverse energy vector and the high-$p_T$ jets: $\Delta\phi$(missing $E_T$, $E_T$(jet)) > 17°.

2.5 Photons and Calibration of the Electromagnetic Calorimeter

The first data recorded at the LHC experiments will be used to establish the calibration and uniformity of the response of the electromagnet calorimeters. At first, the calibration can be obtained from the analysis of neutral pions. With increasing luminosity, events with Z bosons decaying into an electron and a positron will become more important, as they open the probed range to higher energies and provide the kinematical constraint from the well known mass of the Z boson. Figure 6 shows the expected improvements in the ECAL uniformity as a function of recorded data for ATLAS. The figure shows that for a relatively small data set of 100 pb$^{-1}$ integrated luminosity, the constant term in the energy resolution can be greatly improved and that the plateau can be reached for > 300 pb$^{-1}$.

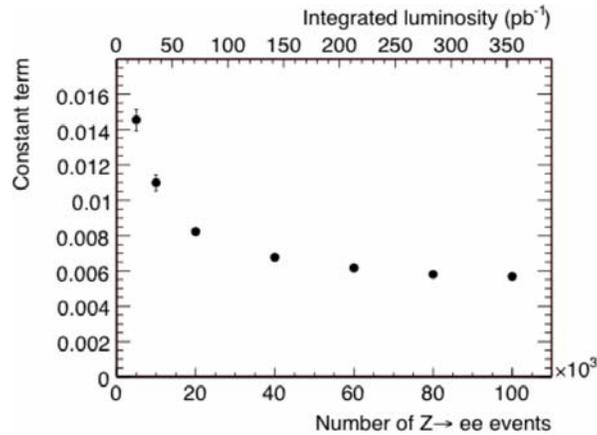

Figure 6: Anticipated improvement in the constant term of the energy resolution, which is related to the overall-uniformity of the electromagnetic calorimeter, as a function of integrated luminosity at ATLAS.

## 3 Conclusions

The commissioning of the CMS detector is in its final phase. More and more subsystems are joining global data taking exercises that use muon signals from cosmic rays for coarse timing, calibration and alignment. Parallel to the commissioning of the detectors, the two multipurpose experiments ATLAS and CMS are preparing the commissioning of the basic physics objects, such as jets, photons and missing transverse energy, with first data. Various methods have been studied with more realistic simulations of the initial calibration and alignment and techniques are developed to establish the sound understanding of the detector response and calibration flows from the early data itself.